\documentclass[reprint,
superscriptaddress,
amsmath,amssymb,
aps,
prl,
]{revtex4-2}
\usepackage{graphicx}
\usepackage{dcolumn}
\usepackage{bm}
\usepackage{dsfont}
\usepackage{physics}
\usepackage{comment}

\begin{document}

\title{General Formulas of the Structure Constants in the $\mathfrak{su}(N)$ Lie Algebra}

\author{Duncan Bossion}
\email{dbossion@ur.rochester.edu}
\affiliation{Department of Chemistry, University of Rochester, Rochester, New York, 14627}
\author{Pengfei Huo}%
 \email{pengfei.huo@rochester.edu}
\affiliation{Department of Chemistry, University of Rochester, Rochester, New York, 14627}
\affiliation{Institute of Optics, University of Rochester, Rochester, New York, 14627}

%
\begin{abstract}
We provide the analytic expressions of the totally symmetric and anti-symmetric structure constants in the $\mathfrak{su}(N)$ Lie algebra. The derivation is based on a relation linking the index of a generator to the indexes of its non-null elements. The closed formulas obtained to compute the values of the structure constants are simple expressions involving those indexes and can be analytically evaluated without any need of the expression of the generators.  We hope that these expressions can be widely used for analytical and computational interest in Physics.
\end{abstract}
\maketitle
The $\mathfrak{su}(N)$ Lie algebra and their corresponding Lie groups are widely used in fundamental physics, particularly in the Standard Model of particle physics~\cite{halzen1984,georgi2000}. The $\mathfrak{su}(2)$ Lie algebra is used describe the spin-$\frac{1}{2}$ system. Its generators, the spin operators, are $\hat{\mathcal S}_j=\frac{\hbar}{2}\sigma_j$ with the Pauli matrices
\begin{align}
    \sigma_1 = \begin{pmatrix}
    0 & 1 \\ 
    1 & 0
    \end{pmatrix},~ \sigma_2 = \begin{pmatrix}
    0 & -i \\ 
    i & 0
    \end{pmatrix},~ \sigma_3 = \begin{pmatrix}
    1 & 0 \\ 
    0 & -1
    \end{pmatrix}. 
\end{align}
The generators of the $\mathfrak{su}(3)$ Lie algebra are proportional to the Gell-Mann $\lambda$ matrices~\cite{gellmann1962} as $\hat{\mathcal S}_j=\frac{\hbar}{2}\lambda_j$, with
\begin{align}
    &\lambda_1 = \begin{pmatrix}
    0 & 1 & 0\\ 
    1 & 0 & 0 \\
    0 & 0 & 0
    \end{pmatrix},~ \lambda_2 = \begin{pmatrix}
    0 & -i & 0 \\ 
    i & 0 & 0 \\
    0 & 0 & 0
    \end{pmatrix},~ \lambda_3 = \begin{pmatrix}
    1 & 0 & 0 \\ 
    0 & -1 & 0 \\
    0 & 0 & 0
    \end{pmatrix}\nonumber\\
    &\lambda_4 = \begin{pmatrix}
    0 & 0 & 1\\ 
    0 & 0 & 0 \\
    1 & 0 & 0
    \end{pmatrix},~ \lambda_5 = \begin{pmatrix}
    0 & 0 & -i \\ 
    0 & 0 & 0 \\
    -i & 0 & 0
    \end{pmatrix},~ \lambda_6 = \begin{pmatrix}
    0 & 0 & 0 \\ 
    0 & 0 & 1 \\
    0 & 1 & 0
    \end{pmatrix}\nonumber\\
    &\hspace{1.2cm}\lambda_7 = \begin{pmatrix}
    0 & 0 & 0\\ 
    0 & 0 & -i \\
    0 & i & 0
    \end{pmatrix},~ \lambda_8 = \frac{1}{\sqrt{3}}\begin{pmatrix}
    1 & 0 & 0 \\ 
    0 & 1 & 0 \\
    0 & 0 & -2
    \end{pmatrix}.
\end{align}
These matrices are used in quantum chromodynamics as an approximate symmetry of the strong interaction between quarks and gluons~\cite{gellmann1962}. There are different manners for obtaining the generators of an algebra, but the most commonly used one in physics is based on a generalization of the Pauli matrices of $\mathfrak{su}(2)$ and of the Gell-Mann matrices~\cite{gellmann1962} of $\mathfrak{su}(3)$, which is what we used in this work. 

The generators of the $\mathfrak{su}(N)$ Lie algebra can be expressed as follows~\cite{Kimura2003,Krammer2008}. There are a total of $N(N-1)/2$ symmetric matrices
\begin{equation}\label{ssym}
    \hat{\mathcal{S}}_{\mathrm{S}_{nm}}=\frac{\hbar}{2}\big(|m\rangle\langle n|+|n\rangle\langle m|\big),
\end{equation}
as well as $N(N-1)/2$ anti-symmetric matrices
\begin{equation}\label{sasym}
    \hat{\mathcal{S}}_{\mathrm{A}_{nm}}=-i\frac{\hbar}{2}\big(|m\rangle\langle n|-|n\rangle\langle m|\big),
\end{equation}
and $N-1$ diagonal matrices (so-called Cartan generators)
\begin{equation}\label{sdiag}
    \hat{\mathcal{S}}_{\mathrm{D}_{n}}=\frac{\hbar}{\sqrt{2n(n-1)}}\Big(\sum_{k=1}^{n-1}|k\rangle\langle k|+(1-n)|n\rangle\langle n|\Big),
\end{equation}
The indexes $\mathrm{S}_{nm}$, $\mathrm{A}_{nm}$ and $\mathrm{D}_{n}$ indicate generators corresponding to the symmetric, anti-symmetric, and diagonal matrices, respectively. The explicit relation between a projection operator $|m\rangle\langle n|$ and the generators can be found in Ref.~\cite{Krammer2008}. Note that these generators are traceless $\mathrm{Tr}[\hat{\mathcal{S}}_{i}]=0$, as well as orthonormal $\mathrm{Tr}[\hat{\mathcal{S}_{i}}\hat{\mathcal{S}_{j}}]=\frac{\hbar^2}{2}\delta_{ij}$. 

These higher dimensional $\mathfrak{su}(N)$ Lie algebra are commonly used in particle physics. For example, the $\mathfrak{su}(6)$ Lie algebra in the quark model~\cite{kokkedee1969}; in the search of a Grand Unification Theory (GUT), $\mathfrak{su}(5)$ has been proposed as the simplest possible version of GUT by Georgi and Glashow~\cite{georgiglashow1974}. In atomic and optical physics~\cite{hioeeberly1981}, as well as in physical chemistry~\cite{meyermiller1979,runesonrichardson2019}, spin analogy is used to map the electronic-nuclear dynamics of open quantum systems involving non-adiabaticity~\cite{meyermiller1979,hiroshi1980,runesonrichardson2019,runesonrichardson2020,bossionhuo2021}. The $\mathfrak{su}(N)$ Lie algebra is also proposed as general mapping between a multi-state Hamiltonian and a classical-like Hamiltonian~\cite{runesonrichardson2020}. In quantum computing, the $\mathfrak{su}(N)$ Lie algebra~\cite{Krammer2008,Zassenhaus1988} is widely used for describing the qudit~\cite{Kais2020}, the state of a $d$-state quantum system.

The crucial quantities to define an algebra are the generators and what govern their commutation and anti-commutation relations. These relations give rise to the structure constants, where the totally anti-symmetric structure constant, $f_{ijk}$, is defined through the commutation relation
\begin{equation}\label{comm}
    [\hat{\mathcal{S}}_i,\hat{\mathcal{S}}_j]=i\hbar\sum_{k=1}^{N^2-1}f_{ijk}\hat{\mathcal{S}}_k,
\end{equation}
which is anti-symmetric under the exchange of any two indexes. The totally symmetric structure constant, $d_{ijk}$, is determined through the anti-commutation relation
\begin{equation}\label{anti-comm}
    \{\hat{\mathcal{S}}_i,\hat{\mathcal{S}}_j\}=\frac{\hbar^2}{N}\delta_{ij}\hat{\mathcal{I}}+\hbar\sum_{k=1}^{N^2-1}d_{ijk}\hat{\mathcal{S}}_k,
\end{equation}
which is symmetric under the exchange of any two indexes. Using the properties in Eq.~\ref{comm} and Eq.~\ref{anti-comm}, as well as the fact that the generators are traceless and orthogonal to each other, one can derive the well-known results~\cite{haber2021}
\begin{equation}\label{fdtrace}
    f_{ijk}=-i\frac{2}{\hbar^3}\mathrm{Tr}\big[[\hat{\mathcal{S}}_i,\hat{\mathcal{S}}_j]\hat{\mathcal{S}}_k\big];~d_{ijk}=\frac{2}{\hbar^3}\mathrm{Tr}\big[\{\hat{\mathcal{S}}_i,\hat{\mathcal{S}}_j\}\hat{\mathcal{S}}_k\big].
\end{equation}
Using the above relation as well as general expressions of the generators (Eq.~\ref{ssym}-\ref{sdiag}) one can compute the numerical values of all the structure constants of the  $\mathfrak{su}(N)$ algebra  through Eq.~\ref{fdtrace}, as has been commonly done in the literature. However, this requires laboriously efforts of different combinations of the $N^2-1$ generators of $\mathfrak{su}(N)$, which remain numerous even when considering the symmetry properties of the structure constants when $N$ is large. Despite the extensive usage and crucial role of these structure constants, to the best of our knowledge~\footnote{On page 106 in Ref.~\cite{pfeifer2003} (Chapter 6), the author suggested that ``In order to determine the structure constants of $\mathfrak{su}(N)$, no closed formulas are known, they have to be calculated by means [Eq.~\ref{fdtrace} in the current letter] of performing matrix multiplications.''}, we are not aware of any analytic expression (closed formulas) of $f_{ijk}$ and $d_{ijk}$.  

In this letter, we derive the analytic expressions of the structure constants in the $\mathfrak{su}(N)$ Lie algebra. The key results are summarized in Eq.~\ref{eq:fijk} for $f_{ijk}$ and Eq.~\ref{eq:dijk} for $d_{ijk}$. We first determine the relation of the indexes of the generators (see Eq.~\ref{ssym}-\ref{sdiag}) $\mathrm{S}_{nm}$, $\mathrm{A}_{nm}$, and $\mathrm{D}_{nm}$ with the label $n$ and $m$. We note that one can use recursive relations to obtain the generators of $\mathfrak{su}(N+1)$ from the generators of $\mathfrak{su}(N)$~\footnote{More specifically, the first $N^2-1$ generators of the $\mathfrak{su}(N+1)$ Lie algebra are directly adapted from those of $\mathfrak{su}(N)$ with adding a $(N+1)$-th row and column of zeros. Then one alternatively introduces the symmetric and anti-symmetric matrices containing the elements from $m=1$ to $m=N$ with $n=N+1$. Lastly, one adds the diagonal matrix of $\hat{\mathcal{S}}_{\mathrm{D}_{N+1}}$ based on the expression in Eq.~\ref{sdiag}. This procedure is apparent from $\mathfrak{su}(2)$ to $\mathfrak{su}(3)$, and the example from $\mathfrak{su}(3)$ to $\mathfrak{su}(4)$ can be found in Chapter 5 of Ref.~\cite{pfeifer2003}}. This helps to determine the indexes of the generators as
\begin{subequations}
    \begin{align}
        \mathrm{S}_{nm} =& n^2+2(m-n)-1,\\
        \mathrm{A}_{nm}=& n^2+2(m-n),\\
        \mathrm{D}_{n}=& n^2-1.
    \end{align}
\end{subequations}
Note that there is no overlap among the indexes as long as the conditions $1\leq m<n\leq N$ holds. Hence, there is a one to one correspondence between the value $i\in\{\mathrm{S}_{nm},\mathrm{A}_{nm},\mathrm{D}_{n}\}$ with indexes $\{n,m\}$,  which helps us to identify the type of generator it labels. This is the first key step to determine closed formulas of structure constants.

\paragraph{\bf The Totally Anti-symmetric Structure Constants $f_{ijk}$.}
The commutation relation between two symmetric generators is
\begin{align}
    &[\hat{\mathcal{S}}_{\mathrm{S}_{nm}},\hat{\mathcal{S}}_{\mathrm{S}_{n'm'}}]=i\hbar\sum_{k=1}^{N^2-1}f_{\mathrm{S}_{nm}\mathrm{S}_{n'm'}k}\hat{\mathcal{S}}_k\\
    &=\frac{\hbar^2}{4}\Big[\delta_{nm'}(|m\rangle\langle n'|-|n'\rangle\langle m|)+\delta_{nn'}(|m\rangle\langle m'|-|m'\rangle\langle m|)\nonumber\\
    &~~~+\delta_{mm'}(|n\rangle\langle n'|-|n'\rangle\langle n|)+\delta_{mn'}(|n\rangle\langle m'|-|m'\rangle\langle n|)\Big]\nonumber\\
    &=i\frac{\hbar}{2}\Big[\delta_{nm'}\hat{\mathcal{S}}_{\mathrm{A}_{n'm}}+\delta_{nn'}(\hat{\mathcal{S}}_{\mathrm{A}_{m'm}}-\hat{\mathcal{S}}_{\mathrm{A}_{mm'}})\nonumber\\
    &~~~+\delta_{mm'}(\hat{\mathcal{S}}_{\mathrm{A}_{n'n}}-\hat{\mathcal{S}}_{\mathrm{A}_{nn'}})-\delta_{mn'}\hat{\mathcal{S}}_{\mathrm{A}_{nm'}}\Big].\nonumber
\end{align}
With constraints from $\delta_{ij}$, we can identify the index of each generator, and hence obtain the non-zero analytic expression of $f_{\mathrm{S}_{nm}\mathrm{S}_{n'm'}k}$, which are summarized in the first line of Eq.~\ref{eq:fijk}.

For a symmetric and an anti-symmetric generator, the commutation relation is
\begin{align}\label{SAcomm}
    &[\hat{\mathcal{S}}_{\mathrm{S}_{nm}},\hat{\mathcal{S}}_{\mathrm{A}_{n'm'}}]=i\hbar\sum_{k=1}^{N^2-1}f_{\mathrm{S}_{nm}\mathrm{A}_{n'm'}k}\hat{\mathcal{S}}_k\\
    &=i\frac{\hbar^2}{4}\Big[\delta_{nn'}(|m\rangle\langle m'|+|m'\rangle\langle m|)-\delta_{nm'}(|m\rangle\langle n'|+|n'\rangle\langle m|)\nonumber\\
    &~~~+\delta_{mn'}(|n\rangle\langle m'|+|m'\rangle\langle n|)-\delta_{mm'}(|n\rangle\langle n'|+|n'\rangle\langle n|)\Big]\nonumber\\
    &=i\frac{\hbar}{2}\Big[\delta_{nn'}(\hat{\mathcal{S}}_{\mathrm{S}_{m'm}}+\hat{\mathcal{S}}_{\mathrm{S}_{mm'}})-\delta_{nm'}\hat{\mathcal{S}}_{\mathrm{S}_{n'm}}\nonumber\\
    &~~~+\delta_{mn'}\hat{\mathcal{S}}_{\mathrm{S}_{nm'}}-\delta_{mm'}(\hat{\mathcal{S}}_{\mathrm{S}_{n'n}}+\hat{\mathcal{S}}_{\mathrm{S}_{nn'}})\nonumber\\
    &~~~+\frac{\hbar}{2}\delta_{mm'}\delta_{nn'}(2|m\rangle\langle m|-2|n\rangle\langle n|)\Big].\nonumber
\end{align}
The first two lines of the last equality in Eq.~\ref{SAcomm} give us directly several structure constants.
The last line of Eq.~\ref{SAcomm} contains diagonal elements, hence we know it will be a combination of diagonal generators. In fact, we can prove (see Supplemental Material~\cite{SI}, Sec I) that
\begin{align}
&i\frac{\hbar^2}{2}(|m\rangle\langle m|-|n\rangle\langle n|)\\
&=i\hbar\Big(\sqrt{\frac{n}{2(n-1)}}\hat{\mathcal{S}}_{\mathrm{D}_n}+\sum_{k>m}^{n-1}\frac{\hat{\mathcal{S}}_{\mathrm{D}_k}}{\sqrt{2k(k-1)}}-\sqrt{\frac{m-1}{2m}}\hat{\mathcal{S}}_{\mathrm{D}_m}\Big).\nonumber
\end{align}
This helps to determine the rest of the structure constant $f_{\mathrm{S}_{nm}\mathrm{A}_{n'm'}k}$ with the expressions documented in Eq.~\ref{eq:fijk}. 
The commutation relations between symmetric and diagonal generators are not required as we already obtained all the non-zero structure constants involving diagonal and symmetric generators (as we know that we cannot obtain a diagonal matrix through the commutator of a symmetric and a diagonal generator).

Between two anti-symmetric generators, the commutation relation is
\begin{align}
    &[\hat{\mathcal{S}}_{\mathrm{A}_{nm}},\hat{\mathcal{S}}_{\mathrm{A}_{n'm'}}]=i\hbar\sum_{k=1}^{N^2-1}f_{\mathrm{A}_{nm}\mathrm{A}_{n'm'}k}\hat{\mathcal{S}}_k\\
    &=\frac{\hbar^2}{4}\Big[\delta_{nm'}(|n'\rangle\langle m|-|m\rangle\langle n'|)+\delta_{nn'}(|m\rangle\langle m'|-|m'\rangle\langle m|)\nonumber\\
    &~~~+\delta_{mm'}(|n\rangle\langle n'|-|n'\rangle\langle n|)+\delta_{mn'}(|m'\rangle\langle n|-|n\rangle\langle m'|)\Big]\nonumber\\
    &=i\frac{\hbar}{2}\Big[-\delta_{nm'}\hat{\mathcal{S}}_{\mathrm{A}_{n'm}}+\delta_{nn'}(\hat{\mathcal{S}}_{\mathrm{A}_{m'm}}-\hat{\mathcal{S}}_{\mathrm{A}_{mm'}})\nonumber\\
    &~~~+\delta_{mm'}(\hat{\mathcal{S}}_{\mathrm{A}_{n'n}}-\hat{\mathcal{S}}_{\mathrm{A}_{nn'}})+\delta_{mn'}\hat{\mathcal{S}}_{\mathrm{A}_{nm'}}\Big],\nonumber
\end{align}
which helps to determine the structure constants involving all anti-symmetric generators (second line of Eq.~\ref{eq:fijk}). The remaining totally anti-symmetric structure constants are computed through the commutator between two diagonal generators, which is $[\hat{\mathcal{S}}_{\mathrm{D}_{n}},\hat{\mathcal{S}}_{\mathrm{D}_{n'}}]=i\hbar\sum_{k=1}^{N^2-1}f_{\mathrm{D}_{n}\mathrm{D}_{n'}k}\hat{\mathcal{S}}_k=0$ (see proof in Supplemental Material, Sec II), indicating a zero value for all $f_{\mathrm{D}_{n}\mathrm{D}_{n'}k}$. This was a known fact, as the diagonal matrices are generators of the Cartan subalgebra of $\mathfrak{su}(N)$, and they commute by definition~\cite{georgi2000}.

To summarize, all of the non-zero totally anti-symmetric structure constants are expressed as follows
\begin{align}\label{eq:fijk}
    &f_{\mathrm{S}_{nm}\mathrm{S}_{kn}\mathrm{A}_{km}}=f_{\mathrm{S}_{nm}\mathrm{S}_{nk}\mathrm{A}_{km}}=f_{\mathrm{S}_{nm}\mathrm{S}_{km}\mathrm{A}_{kn}}=\frac{1}{2},\\
    &f_{\mathrm{A}_{nm}\mathrm{A}_{km}\mathrm{A}_{kn}}=\frac{1}{2},\nonumber\\
    &f_{\mathrm{S}_{nm}\mathrm{A}_{nm}\mathrm{D}_{m}}=-\sqrt{\frac{m-1}{2m}},~~~f_{\mathrm{S}_{nm}\mathrm{A}_{nm}\mathrm{D}_{n}}=\sqrt{\frac{n}{2(n-1)}},\nonumber\\
    &~~~~~~~~~f_{\mathrm{S}_{nm}\mathrm{A}_{nm}\mathrm{D}_{k}}=\sqrt{\frac{1}{2k(k-1)}},~m<k<n.\nonumber
\end{align}
One of the interesting usages of these expressions is the construction of an adjoint representation of the $\mathfrak{su}(N)$ Lie algebra, which is a defining representation of SO($N^2-1$), whose generators $\hat{\mathcal T}_i$ are $(N^2-1)\times(N^2-1)$ matrices. The $(jk)$-th matrix element of $\hat{\mathcal T}_i$ is $[\hat{\mathcal T}_i]_{jk}=-if_{ijk}$. Thus, $\hat{\mathcal T}_i$ are anti-symmetric, non-diagonal and traceless, and we provide analytic expressions to obtain them through the totally anti-symmetric structure constants expressions.

\paragraph{\bf Totally Symmetric Structure Constants $d_{ijk}$.}
The anti-commutation relation between two symmetric generators is
\begin{align}\label{dSS}
    &\{\hat{\mathcal{S}}_{\mathrm{S}_{nm}},\hat{\mathcal{S}}_{\mathrm{S}_{n'm'}}\}=\frac{\hbar^2}{N}\delta_{\mathrm{S}_{nm}\mathrm{S}_{n'm'}}\hat{\mathcal{I}}+\hbar\sum_{k=1}^{N^2-1}d_{\mathrm{S}_{nm}\mathrm{S}_{n'm'}k}\hat{\mathcal{S}}_k\nonumber\\
    &=\frac{\hbar^2}{4}\Big[\delta_{nm'}(|m\rangle\langle n'|+|n'\rangle\langle m|)+\delta_{nn'}(|m\rangle\langle m'|+|m'\rangle\langle m|)\nonumber\\
    &~~~+\delta_{mm'}(|n\rangle\langle n'|+|n'\rangle\langle n|)+\delta_{mn'}(|n\rangle\langle m'|+|m'\rangle\langle n|)\Big]\nonumber\\
    &=\frac{\hbar}{2}\Big[\delta_{nm'}\hat{\mathcal{S}}_{\mathrm{S}_{n'm}}+\delta_{nn'}(\hat{\mathcal{S}}_{\mathrm{S}_{m'm}}+\hat{\mathcal{S}}_{\mathrm{S}_{mm'}})\nonumber\\
    &~~~+\delta_{mm'}(\hat{\mathcal{S}}_{\mathrm{S}_{n'n}}+\hat{\mathcal{S}}_{\mathrm{S}_{nn'}})+\delta_{mn'}\hat{\mathcal{S}}_{\mathrm{S}_{nm'}}\nonumber\\
    &~~~+\delta_{nn'}\delta_{mm'}\frac{\hbar}{2}(2|m\rangle\langle m|+2|n\rangle\langle n|)\Big].
\end{align}
We know that the last line of Eq.~\ref{dSS} only involves diagonal matrices. In fact, we can prove that (see Supplemental Material, Sec III) 
\begin{align}\label{SS-diag}
    &\frac{\hbar^2}{2}(|m\rangle\langle m|+|n\rangle\langle n|)=\frac{\hbar^2}{N}\hat{\mathcal{I}}\\
    &~~~~~~~~+\hbar\Big(\sum_{k>n}^N\sqrt{\frac{2}{k(k-1)}}\hat{\mathcal{S}}_{\mathrm{D}_k}+\frac{2-n}{\sqrt{2n(n-1)}}\hat{\mathcal{S}}_{\mathrm{D}_n}\nonumber\\
    &~~~~~~~~+\sum_{k=m+1}^{n-1}\sqrt{\frac{1}{2k(k-1)}}\hat{\mathcal{S}}_{\mathrm{D}_k}-\sqrt{\frac{m-1}{2m}}\hat{\mathcal{S}}_{\mathrm{D}_m}\Big).\nonumber
\end{align}
Thus, we can extract all the non-zero $d_{\mathrm{S}_{nm}\mathrm{S}_{n'm'}k}$, with the expressions summarized in Eq.~\ref{eq:dijk}.

Between a symmetric and an anti-symmetric generator, the anti-commutation relation reads
\begin{align}\label{SA-anti}
    &\{\hat{\mathcal{S}}_{\mathrm{S}_{nm}},\hat{\mathcal{S}}_{\mathrm{A}_{n'm'}}\}=\hbar\sum_{k=1}^{N^2-1}d_{\mathrm{S}_{nm}\mathrm{A}_{n'm'}k}\hat{\mathcal{S}}_k\\
    &=i\frac{\hbar^2}{4}\Big[\delta_{nm'}(|n'\rangle\langle m|-|m\rangle\langle n'|)+\delta_{nn'}(|m\rangle\langle m'|-|m'\rangle\langle m|)\nonumber\\
    &~~~+\delta_{mm'}(|n'\rangle\langle n|-|n\rangle\langle n'|)+\delta_{mn'}(|n\rangle\langle m'|-|m'\rangle\langle n|)\Big]\nonumber\\
    &=\frac{\hbar}{2}\Big[\delta_{nm'}\hat{\mathcal{S}}_{\mathrm{A}_{n'm}}+\delta_{nn'}(\hat{\mathcal{S}}_{\mathrm{A}_{mm'}}-\hat{\mathcal{S}}_{\mathrm{A}_{m'm}})\nonumber\\
    &~~~+\delta_{mm'}(\hat{\mathcal{S}}_{\mathrm{A}_{n'n}}-\hat{\mathcal{S}}_{\mathrm{A}_{nn'}})+\delta_{mn'}\hat{\mathcal{S}}_{\mathrm{A}_{nm'}}\Big],\nonumber
\end{align}
from which one can extract $d_{\mathrm{S}_{nm}\mathrm{A}_{n'm'}k}$. Note that based on Eq.~\ref{SA-anti}, there is no diagonal component $\hat{\mathcal S}_\mathrm{D}$, thus all the $d_{\mathrm{S}\mathrm{A}\mathrm{D}}=0$. 
The anti-commutator between symmetric and diagonal generators is not necessary as we already obtained the structure constants involving those generators by permutation~\footnote{Note that $d_{\mathrm{S}\mathrm{D}\mathrm{D'}}$ will be same as $d_{\mathrm{D}\mathrm{D'}\mathrm{S}}$, and $d_{\mathrm{S}\mathrm{D}\mathrm{S'}}$ will be same as $d_{\mathrm{S}\mathrm{S'}\mathrm{D}}$, and $d_{\mathrm{S}\mathrm{D}\mathrm{A}}$ will be same as $d_{\mathrm{S}\mathrm{A}\mathrm{D}}$}.

Computing the anti-commutators between two anti-symmetric generators gives
\begin{align}\label{eq:AA-anti}
    &\{\hat{\mathcal{S}}_{\mathrm{A}_{nm}},\hat{\mathcal{S}}_{\mathrm{A}_{n'm'}}\}=\frac{\hbar^2}{N}\delta_{\mathrm{A}_{nm}\mathrm{A}_{n'm'}}\hat{\mathcal{I}}+\hbar\sum_{k=1}^{N^2-1}d_{\mathrm{A}_{nm}\mathrm{A}_{n'm'}k}\hat{\mathcal{S}}_k\nonumber\\
    &=\frac{\hbar^2}{4}\Big[\delta_{nn'}(|m\rangle\langle m'|+|m'\rangle\langle m|)-\delta_{nm'}(|m\rangle\langle n'|+|n'\rangle\langle m|)\nonumber\\
    &~~~+\delta_{mm'}(|n\rangle\langle n'|+|n'\rangle\langle n|)-\delta_{mn'}(|n\rangle\langle m'|+|m'\rangle\langle n|)\Big]\nonumber\\
    &=\frac{\hbar}{2}\Big[\delta_{nn'}(\hat{\mathcal{S}}_{\mathrm{S}_{m'm}}+\hat{\mathcal{S}}_{\mathrm{S}_{mm'}})-\delta_{nm'}\hat{\mathcal{S}}_{\mathrm{S}_{n'm}}\nonumber\\
    &~~~+\delta_{mm'}(\hat{\mathcal{S}}_{\mathrm{S}_{n'n}}+\hat{\mathcal{S}}_{\mathrm{S}_{nn'}})-\delta_{mn'}\hat{\mathcal{S}}_{\mathrm{S}_{nm'}}\nonumber\\
    &~~~+\delta_{nn'}\delta_{mm'}\frac{\hbar}{2}(2|m\rangle\langle m|+2|n\rangle\langle n|)\Big],
\end{align}
where we recognize that the last line of Eq.~\ref{eq:AA-anti} is identical to the last line of Eq.~\ref{dSS}, which can be expressed as generators in Eq.~\ref{eq:AA-anti}. We do not need to compute the anti-commutator between an asymmetric and a diagonal generator as we already have the result by permutation from Eq.~\ref{eq:AA-anti} (and Eqs.~\ref{SA-anti} indicates $d_\mathrm{SAD}=0$).

The remaining $d_{ijk}$ values are obtained through the anti-commutator between two diagonal generators
\begin{align}\label{dd-anti}
    &\{\hat{\mathcal{S}}_{\mathrm{D}_{n}},\hat{\mathcal{S}}_{\mathrm{D}_{n'}}\}=\frac{\hbar^2}{N}\delta_{\mathrm{D}_{n}\mathrm{D}_{n'}}\hat{\mathcal{I}}+\hbar\sum_{k=1}^{N^2-1}d_{\mathrm{D}_{n}\mathrm{D}_{n'}k}\hat{\mathcal{S}}_k\\
    &=\frac{\hbar^2}{\sqrt{2n(n-1)2n'(n'-1)}}\Big[\sum_{k=1}^{n-1}\delta_{kk'}(|k\rangle\langle k'|+|k'\rangle\langle k|)\nonumber\\
    &~~~+\delta_{kn'}(1-n')(|k\rangle\langle n'|+|n'\rangle\langle k|)\nonumber\\
    &~~~+\delta_{nk'}(1-n)(|n\rangle\langle k'|+|k'\rangle\langle n|)\nonumber\\
    &~~~+\delta_{nn'}(1-n)(1-n')(|n\rangle\langle n'|+|n'\rangle\langle n|)\Big]\nonumber\\
    &=\frac{\hbar}{\sqrt{2n(n-1)}}2\delta_{kn'}\hat{\mathcal{S}}_{\mathrm{D}_{n'}}+\frac{\hbar}{\sqrt{2n'(n'-1)}}2\delta_{nk'}\hat{\mathcal{S}}_{\mathrm{D}_{n}}\nonumber\\
    &~~~+\delta_{nn'}\frac{\hbar^2}{n(n-1)}(\sum_{k=1}^{n-1}|k\rangle\langle k|+(1-n)^2|n\rangle\langle n|).\nonumber
\end{align}
One can see that only diagonal matrices are involved in the last line of Eq.~\ref{dd-anti} and there is no off-diagonal element. We can prove that (see Supplemental Material, Sec IV)
\begin{align}
    &\frac{\hbar^2}{n(n-1)}(\sum_{k=1}^{n-1}|k\rangle\langle k|+(1-n)^2|n\rangle\langle n|)=\frac{\hbar^2}{N}\hat{\mathcal{I}}\\
    &~~~~~~+\hbar\big(\sum_{k>n}^N\sqrt{\frac{2}{k(k-1)}}\hat{\mathcal{S}}_{\mathrm{D}_k}+(2-n)\sqrt{\frac{2}{n(n-1)}}\hat{\mathcal{S}}_{\mathrm{D}_n}\big),\nonumber
\end{align}
which helps to determine all $d_{\mathrm{D}_{n}\mathrm{D}_{n'}k}$.
 
We summarize all the non-zero totally symmetric structure constants as follows
\begin{align}\label{eq:dijk}
    &d_{\mathrm{S}_{nm}\mathrm{S}_{kn}\mathrm{S}_{km}}=d_{\mathrm{S}_{nm}\mathrm{A}_{kn}\mathrm{A}_{km}}=d_{\mathrm{S}_{nm}\mathrm{A}_{mk}\mathrm{A}_{nk}}=\frac{1}{2},\\
    &d_{\mathrm{S}_{nm}\mathrm{A}_{nk}\mathrm{A}_{km}}=-\frac{1}{2},\nonumber\\
    &d_{\mathrm{S}_{nm}\mathrm{S}_{nm}\mathrm{D}_{m}}=d_{\mathrm{A}_{nm}\mathrm{A}_{nm}\mathrm{D}_{m}}=-\sqrt{\frac{m-1}{2m}},\nonumber\\
    &d_{\mathrm{S}_{nm}\mathrm{S}_{nm}\mathrm{D}_{k}}=d_{\mathrm{A}_{nm}\mathrm{A}_{nm}\mathrm{D}_{k}}=\sqrt{\frac{1}{2k(k-1)}},~m<k<n,\nonumber\\
    &d_{\mathrm{S}_{nm}\mathrm{S}_{nm}\mathrm{D}_{n}}=d_{\mathrm{A}_{nm}\mathrm{A}_{nm}\mathrm{D}_{n}}=\frac{2-n}{\sqrt{2n(n-1)}},\nonumber\\
    &d_{\mathrm{S}_{nm}\mathrm{S}_{nm}\mathrm{D}_{k}}=d_{\mathrm{A}_{nm}\mathrm{A}_{nm}\mathrm{D}_{k}}=\sqrt{\frac{2}{k(k-1)}},~n<k,\nonumber\\
    &d_{\mathrm{D}_{n}\mathrm{D}_{k}\mathrm{D}_{k}}=\sqrt{\frac{2}{n(n-1)}},~k<n,\nonumber\\
    &d_{\mathrm{D}_{n}\mathrm{D}_{n}\mathrm{D}_{n}}=(2-n)\sqrt{\frac{2}{n(n-1)}}.\nonumber
\end{align}
In a recent work on deriving the quantum Liouvillian of coupled electronic-nuclear DOFs based upon the $\mathfrak{su}(N)$ representation (see Appendix F in Ref.~\cite{runesonrichardson2020}), these $d_{ijk}$ are explicitly present in the equation of motion. Having the above analytic expressions will facilitate future theoretical developments.

\paragraph{\bf Mapping Hamiltonian using the $ \mathfrak{su}(N)$ Lie Algebra.} The SU($2$) representation of the Lie algebra (spin-$\frac{1}{2}$ analogy) is often used in quantum dynamics to study systems with two states~\cite{meyermiller1979, runesonrichardson2019, bossionhuo2021}. For a two level system with the Hamiltonian $\hat{H} = H_0\hat{\mathcal{I}} + \frac{1}{\hbar}\mathbf{H}\cdot\hat{\mathbf{S}}=H_0\hat{\mathcal{I}} + \sum_{k}\frac{1}{\hbar}({H}_{k}\cdot\hat{S}_{k})=H_0\hat{\mathcal{I}} + \frac{1}{\hbar}[2 \mathcal{R}(V_{12})\cdot \hat{S}_{x} + 2 \mathcal{I}(V_{12})\cdot \hat{S}_{y}+(V_{11}-V_{22})\cdot \hat{S}_{z}]$,
it can be shown (through the Heisenberg equations of motion (EOMs)) that
\begin{equation}\label{bloch}
    \frac{\mathrm{d}}{\mathrm{d}t}{\mathcal{S}}_i=\frac{1}{\hbar}\sum_{j,k}^3\varepsilon_{ijk}H_j{\mathcal{S}}_k,
\end{equation}
where ${\mathcal{S}}_{i}=\mathrm{Tr}[\hat{\rho}\hat{\mathcal{S}}_{i}]$ is the expectation value of $\hat{\mathcal{S}}_i$, $\hat{\rho}$ being the density operator, and $\varepsilon_{ijk}$ the Levi-Civita tensor which is the two-dimension totally anti-symmetric structure constant $f_{ijk}$. This is equivalent to the precession of spin of a spin-$\frac{1}{2}$ system around a magnetic field $\bf {M}=\bf {H}$, which is a well-known result (eg, Page 424 of Ref.~\cite{Cohen-Tannoudji}). Those EOMs exactly obey the time-dependent Schr\"odinger equation (TDSE), $\dot{c}_i=-i\sum_{j=1}^{2}V_{ij}c_j$. More specifically, with an arbitrary state defined as $|\Psi\rangle=c_1|1\rangle+c_2|2\rangle$, by using the transformation $S_x=\Re\{c_1^*c_2\}$, $S_y=\Im\{c_1^*c_2\}$ and $S_z=\frac{1}{2}(|c_1|^2-|c_2|^2)$, one can show that Eq.~\ref{bloch} is equivalent to TDSE.

For a system with $N$ states, one can use the $\mathfrak{su}(N)$ Lie algebra for the spin analogy~\cite{meyermiller1979,hioeeberly1981,runesonrichardson2020}, describing the precession of the $N$-states system~\footnote{Note that in particular, }. This $\mathfrak{su}(N)$ analogy is based on a reformulation of the Hamiltonian $\hat{H}=H_{0}\hat{\mathcal{I}}+\hat{V}_e(\hat{R})$ with the generators of $\mathfrak{su}(N)$ as follows~\cite{hioeeberly1981,runesonrichardson2020}
\begin{equation}
\hat{H}=\mathcal{H}_0\hat{\mathcal{I}}+\frac{1}{\hbar}\sum_{k=1}^{N^2-1}\mathcal{H}_k\hat{\mathcal{S}}_{k},
\end{equation}
where $\mathcal{H}_k=\frac{1}{\hbar}\mathrm{Tr}[\hat{H}\hat{\mathcal S}_k]$. Similarly, for the density matrix~\cite{hioeeberly1981} $\hat{\rho}=\frac{1}{N}\hat{\mathcal I}+\frac{1}{2}\sum_{k=1}^{N^2-1}\mathcal{S}_{k}\hat{\mathcal{S}}_{k}$ where $\mathcal{S}_{k}=\mathrm{Tr}[\hat{\rho}\hat{\mathcal{S}}_{k}]$.
Plugging the $\mathfrak{su}(N)$ generator expression of $\hat{H}$ and $\hat{\rho}$ into the quantum Liouville equation $i\hbar {\partial\hat{\rho}}/{\partial t} = [\hat{H}, \hat{\rho}]$, one arrives at the following equation~\footnote{This means that } which can be viewed as the  generalization of the spin precession~\cite{hioeeberly1981}
\begin{equation}\label{dSdt}
    \frac{\mathrm{d}}{\mathrm{d}t}{\mathcal{S}}_i=\frac{1}{\hbar}\sum_{j,k=1}^{N^2-1}f_{ijk}H_j{\mathcal{S}}_k.
\end{equation}
where $f_{ijk}$ is the totally anti-symmetric structure constant of $\mathfrak{su}(N)$. 
For an arbitrary state defined as $|\Psi\rangle=\sum_{k=1}^Nc_k|k\rangle$, using the transformation
\begin{align}
S_{\mathrm{S}_{nm}}&=\Re\{c_m^*c_n\},~~~ S_{\mathrm{A}_{nm}}=\Im\{c_m^*c_n\},\\ S_{\mathrm{D}_n}&=\sum_{k=1}^{n-1}\frac{1}{\sqrt{2n(n-1)}}|c_k|^2-\sqrt{\frac{n-1}{2n}}|c_n|^2,
\end{align}
and the analytic expressions of the $f_{ijk}$, one can show that Eq.~\ref{dSdt} is equivalent to the TDSE . Hence Eq.~\ref{dSdt} now has a closed formula.

\paragraph{\bf Conclusion.}
In this letter, we provide the analytic expressions of the 
totally symmetric and totally anti-symmetric structure constants for the $\mathfrak{su}(N)$ Lie algebra. We hope that these expressions can be widely used for analytical and computational interest in Physics, as they are valid for any dimension $N$ of $\mathfrak{su}(N)$ Lie algebra without the need to explicitly compute the commutation and anti-commutation relations or use any generator. The structure constants bear important information on the algebra they belong to, and the possibility to obtain those constants with simple relations can bring insight into the high dimensional $\mathfrak{su}(N)$ Lie algebra, which might be challenging otherwise. 


\section*{acknowledgments}
This work was supported by the National Science Foundation CAREER Award under Grant No. CHE-1845747.

%
%
%
%
%
\end{document}